\documentclass[12pt,preprint]{aastex}

\newcommand{\Ha}{H$\alpha$}

\newcommand{\um}{$\mu$m}

\slugcomment{Submitted to the Astronomical Journal}

\shorttitle{Molecular Knots in Helix}
\shortauthors{Meixner et al.}

\begin{document}


\title{The Multitude of Molecular Hydrogen Knots in the Helix Nebula\footnote{
Based in part on observations with the NASA/ESA Hubble Space Telescope,
obtained at the Space Telescope Science Institute, which is operated by
the Association of Universities for Research in Astronomy, Inc., under
NASA Contract No.~NAS 5-26555.}}


\author{Margaret Meixner and Peter McCullough}
\affil{Space Telescope Science Institute, 3700 San Martin Drive, Baltimore,
MD 21218}
\email{meixner@stsci.edu,pmcc@stsci.edu}
\author{Joel Hartman} \affil{Harvard-Smithsonian Center for Astrophysics, 60 Garden St., Cambridge, MA
02138}
\email{jhartman@cfa.harvard.edu}
\author{Minho Son} \affil{Department of Physics and Astronomy, Johns Hopkins University,
Baltimore, MD 21218}
\email{mhson@pha.jhu.edu}
\and

\author{Angela Speck}
\affil{Dept. of Physics, University of Missouri, Columbia, MO,}
\email{speckan@missouri.edu}

\begin{abstract}
We present HST/NICMOS imaging of the H$_2$ 2.12~$\mu$m emission in 5~fields in the Helix Nebula ranging in radial distance from 250--450\arcsec~from the central star. The images reveal arcuate structures with their apexes pointing towards the central star.   These molecular hydrogen knots are most highly structured in the fields closest to the central star and become increasingly less structured with increasing radius. Comparison of these images with comparable resolution ground based images reveals that the molecular gas is more highly clumped than the ionized gas line tracers.  From our images,  we determine an average number density of knots in the molecular gas ranging from 162 knots/arcmin$^2$ in the denser regions to 18~knots/arcmin$^2$ in the lower density outer regions.  The decreasing number density of H$_2$ knots in the outer regions creates a lower filling factor of neutral and molecular gas emission in the radio observations of CO and HI  and may  explain why these outer regions, where we clearly detect H$_2$ 2.12~$\mu$m,  fall below the detection limit of the radio observations. Using this new number density, we estimate that the total number of knots in the Helix to be $\sim$23,000 which is a factor of 6.5 larger than previous estimates.  The total neutral gas  mass in the Helix  is 0.35~M$_\odot$ assuming a mass of 
$\sim$$1.5\times10^{-5}$~M$_\odot$   for the individual knots.  The H$_2$ emission structure of  the entire Helix nebula  supports the recent interpretation of the Helix as a nearly  pole-on  poly-polar planetary nebula. The H$_2$  intensity,  5--9$\times 10^{-5}$  erg~s$^{-1}$ cm$^{-2}$ sr$^{-1}$, remains relatively constant with projected distance from the central star suggesting a heating mechanism for the molecular gas that is distributed almost uniformly in the knots throughout the nebula. The temperature and H$_2$ 2.12~$\mu$m  intensity of the knots can be approximately explained by  photodissociation  regions (PDRs) in the individual knots; however, theoretical PDR models  of PN under-predict the intensities of some knots by a factor of 10. The brightest H$_2$ emission ($\sim 3\times 10^{-4}$  erg~s$^{-1}$ cm$^{-2}$ sr$^{-1}$) may be enhanced by a larger than unity area filling factor of  H$_2$  knots or may be an individual H$_2$ knot exposed to direct starlight causing rapid photoevaporation compared with the more embedded knots of the disk.
\end{abstract}


\section{INTRODUCTION}
Approximately 50 planetary nebulae (PNs) are presently known to have ``small scale" heterogeneities located inside or outside the main ionized nebulae \citep{goncalves01}.  Cometary knots are a subcategory of small scale structures found commonly in nearby, evolved PNsPNs \citep{odell02}.  Because of its closest proximity (213 parsecs \cite{harris97}),  the Helix Nebula (NGC~7293) is the best case to study the structure and excitation conditions of cometary knots.  The nature of the cometary knots in the Helix was  first established by \cite{meaburn92}.  The detailed structure  of the cometary knots   has been resolved in ionized gas lines in the optical by \cite{odell96} with further detailed analysis by \cite{burkert98}, and \cite{odell00}.  The emerging optical picture  of the cometary knots reveals that they are neutral gas condensations that appear as comet like structures with rims bright in H$\alpha$ and tails that appear as shadows in [OIII]  and that point away from the central star.   The rim of low-excitation ionize gas has a steep temperature gradient  indicating that the knots are photo-evaporating and that ionization fronts are advancing into the knots \citep{odell00}.  A recent anlaysis of knots over the whole Helix nebula  by \cite{odell04} revealed a new 3-D picture for the main ring of the Helix:  it is composed of a disk structure and an outer ring tilted almost perpendicularly with respect to the disk.  Within each of these components, they observed a similar, progressive evolution in the structure of the knots.  The knots closest to the central star and clearly inside of the ionization front were elegantly carved with the brightest rims.  The knots furthest from the central star appeared slightly more amorphous in their structure with less well defined rims.  The culmination of these optical observations appear to support the theory that these knots were  initially formed earlier by instabilities at the ionization front  or perhaps by the interaction of the fast stellar wind  and then have been sculpted by interaction with the harsh radiation field of the central star \citep{capriotti73}.  

In contrast to the high angular resolution ($\sim$0.01\arcsec) optical studies of the ionized gas lines in the cometary knots, the molecular gas observations  have  had lower angular resolution (4\arcsec--41\arcsec) and sensitivity making it difficult to determine the detailed structure and excitation of the main gas component of the cometary knots.   These low resolution studies have revealed that the Helix has retained a significant amount of molecular gas  \citep{young99,huggins86,speck02}  and that the molecular gas appears to be very clumpy and is probably confined to  cometary knot structures \citep{speck02, huggins02}.     The only  detailed study  of  an isolated cometary knot, which is  close to the central star,  shows no evidence for  large velocities in the molecular gas, ruling out a stellar wind shaping the knot,  and reveals
a stratified structure for the  ionized and molecular gas emissions that is expected in a photodissociation region (PDR) \citep{huggins02}.  However, since recent optical studies show an evolution of the knot structure with radial distance from the central star \citep{odell04},  it is not clear that this single knot study is representative of all the knots in the nebula. In order  to determine the structure and excitation of the H$_2$ emission in the cometary knots at comparable resolution to optical images across the Helix,  we pursued high angular resolution ($\sim$0.2 \arcsec) NICMOS/NIC3 F212N H$_2$ images at several locations in the nebula, in parallel with the HST/ACS program recently published by \cite{odell04}.   

The remainder of this paper is organized as follows.  In section~2, we  report the observation and data processing procedures.  In 
section~3, we discuss the major observational results  and how these relate to the optical ionized gas line emissions imaged by 
\cite{odell04}.  In section~4, we interpret the observations in the context of current understanding of the Helix's 3-D structure and discuss the number density, mass, evolution and excitation of the knots as revealed by our H$_2$ images.  We summarize our conclusions in section~5.

\section{OBSERVATIONS}

\begin{figure}
\epsscale{.5}
\plotone{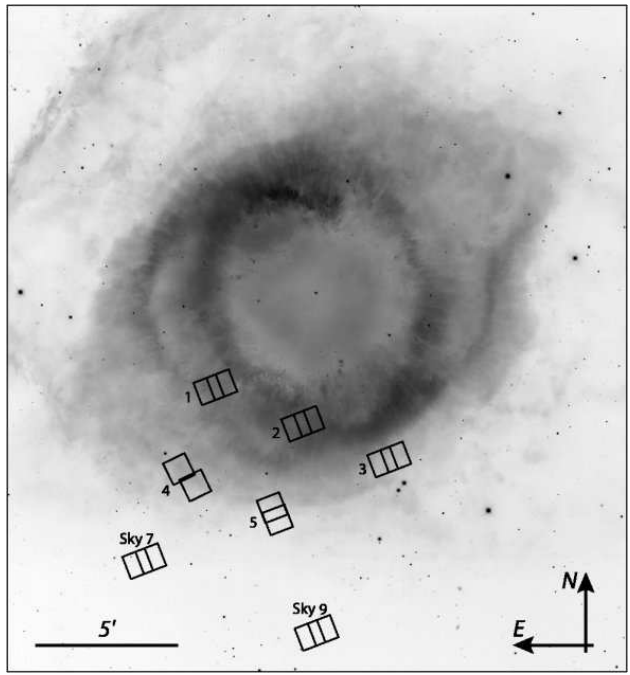}
\caption{This negative picture of the Helix from \cite{odell04} shows the optical structure on a large scale. The NICMOS field positions are overlaid as black rectangles.\label{helixmap}}
\end{figure}

The Hubble Helix project (GO program 9700; PI: M.~Meixner) imaged the Helix nebula during the 2002 Leonids meteor shower that presented a risk to the HST. The imaging involved a 9-panel mosaic of the Helix using the ACS WFC instrument  in the F658N filter (transmitting equally well both the \Ha\ 6563~\AA\ and [N~II] 6584~\AA\ lines) and the F502N filter (dominated by the [O~III] 5007~\AA\ line).  In parallel with the ACS imaging, we used  NICMOS \citep{thompson98} to image 7 of the possible 9~field positions,  5 of which landed on the nebula (positions~1, 2, 3, 4 and~5) and~2 of which were off the nebula (positions~7 and~9) and used for background measurements for the 5~fields on the nebula.  Figure~\ref{helixmap} shows the location of these fields on  the Helix and the RA and Dec of  the  field centers for field positions~1, 2, 3, 4, and~5 are listed in Table~\ref{loctab}.  These parallel NICMOS field positions had insignificant overlap with the the ACS images. Because we wanted maximum field of view and our target was a diffuse nebula,   we used the NIC3 camera, $0\farcs2$  pixel$^{-1}$,   with  the  F212N filter  to image  the  H$_2$ 2.12~\um\ line emission in the nebula.  For field positions~1 and~2, half the time was spent in the Pa$\alpha$ filter F187N that is sufficiently low signal-to-noise as to be useless and is not discussed further.   For each field position,  the two dither positions for ACS resulted in two slightly overlapping NICMOS/NIC3 images.  The NIC3 MULTIACCUM, FAST readout mode was used.   The Hubble Helix project and its results \citep{mccullough02}  immediately went into the public domain.  The ACS images were analyzed in combination with ground based CTIO images in similar filters and have been published by \cite{odell04}.
In this work we analyze and discuss the NICMOS H$_2$ 2.12~\um\ emission and its relation to the ionized gas at high spatial resolution.
 
The NICMOS/NIC3 images were reduced and calibrated using the standard set of NICMOS calibration programs provided in the latest version (Version~3.1) of IRAF/STSDAS\footnote{STSDAS is a product of the Space Telescope Science Institute, which is operated by AURA for NASA}. The CALNICA calibration routines in STSDAS perform zero-read signal correction, bias subtraction, dark subtraction, detector non-linearity correction, flat-field correction, and flux calibration.  The pedestal effect was removed by first manually inserting the 
STSDAS task {\sl biaseq} in the middle of the CALNICA processes (before flat-fielding) and then employing the STADAS task {\sl pedsub} after the CALNICA processes.  Cosmic rays were identified and replaced by the median filtered pixel value.

The four dither positions for field positions~7 and~9 were combined to make a ``sky" image that is completely attributed to the telescope emission.  This ``sky"  was subtracted from each of the dither frames for field positions 1--5 resulting in a H$_2$ dominated emission frame.    The continuum emission from the Helix nebula in the F212N filter is negligible in the sky-subtracted images as demonstrated by \cite{speck02}.   The two dither positions for each field position were combined using drizzle which magnifies the images by a factor of~2.  The final drizzled images  have a plate scale of $0\farcs10$  pixel$^{-1}$ and have been rotated so that north is up and east is to the left. Total integration times  for the final, drizzled F212N images ranged from 768 seconds  for field positions~1 and~2  to 1792 seconds  for field position~3.    Figures~\ref{nicpos1}, \ref{nicpos2}, \ref{nicpos3}, \ref{nicpos4} and \ref{nicpos5} show the final NICMOS images for field positions~1, 2, 3, 4 and~5, respectively,  in comparison with the optical emission line images from \cite{odell04}.  These final NICMOS images, in units of count rate per pixel, are available at the MAST web site under the Hubble Helix project.

In order to discern the relative distribution of the H$_2$ line emission with the ionized gas line tracers, we compare our results with comparable resolution optical images.  The overlap between the NICMOS fields and the ACS fields is insignificant.  Fortunately, \cite{odell04} presented ground based CTIO images of H$\alpha$/[NII], [OIII], H$\beta$ and [SII] of the entire Helix Nebula at comparable resolution to our NICMOS images. We registered these CTIO images to the WCS of the NICMOS  images.  The initial comparison, using just the absolute coordinates of the NICMOS and CTIO images, permitted a close enough alignment to identify at least 1~star in common between the NICMOS and CTIO images that was used for translational alignment.   The four CTIO images had 6--7~stars in common with each other and our detailed comparison revealed small rotational errors between them up to 0.08\arcdeg\ relative angular rotations.  Using the star in common with the NICMOS image as the ``origin," we improved the relative rotational alignment  of the CTIO images to better than 0.005\arcdeg.  The CTIO images were then translated  to the NICMOS WCS position, by aligning the star in common using the task  ``register'' in IRAF.  Only the part of the CTIO images that overlap with the NICMOS field positions are shown. 

For each field position, we selected a prominent H$_2$ knot, labeled  with a cross in the figures,  for  surface brightness measurements 
of the H$_2$ emission and made a cross-cut along the radial direction from the central star in order to quantify the relation between H$_2$ emission and distance from the central star.  These positions and cross-cuts are labeled on Figs.~\ref{nicpos1}--\ref{nicpos5} and  the  locations of the H$_2$ knots with respect to the field centers are listed in Table~\ref{loctab}.  Table~\ref{surbrtab} lists the surface brightness measurements for  H$_2$ and the optical line tracers at the knot position.   The  average surface brightness of the H$_2$ knot emission (DN/pixel) was determined for a circular aperture enclosing the brightest part of the H$_2$ emission ($\sim$3~pixel radius centered on the +).  We then converted this average into physical units by multiplying by  $2.44929\times10^{-18}$  erg~s$^{-1}$ cm$^{-2}$ \AA$^{-1}$ DN$^{-1}$ (DN~= counts per second), which is the  photometry conversion keyword derived for the NICMOS, NIC3 camera, filter F212N for the 77.1~K detector, appropriate for observations taken after January 2002.  Finally, to arrive at the units in Table~\ref{surbrtab}, we mulitplied by 212.1~\AA\ which is the filter bandwidth of the F212N filter,  and divided by $2.2350443\times10^{-13}$ sr which is the solid angle of the $0\farcs10$ pseudo-pixel.  For each of the measured H$_2$ knots in the NICMOS fields,  we measured the surface brightness of the  ionized gas emission lines, H$\alpha$[NII], [OIII], H$\beta$ and [SII], in the same way and list the results in Table~\ref{surbrtab}.  A conversion factor of $1.66\times10^{10}$ was used to convert ADUs/pixel to photons s$^{-1}$ cm$^{-2}$ sr$^{-1}$ for all the CTIO images and then each image was multiplied by its photon energy, $h c \over \lambda$  to determine the surface brightness in units of  erg~s$^{-1}$ cm$^{-2}$ sr$^{-1}$, the same as the NICMOS H$_2$ line measurements.  For the cross-cuts shown in Figure~\ref{nicprof}, we applied the same conversion factors.

\section{Results}

\begin{figure}
\plotone{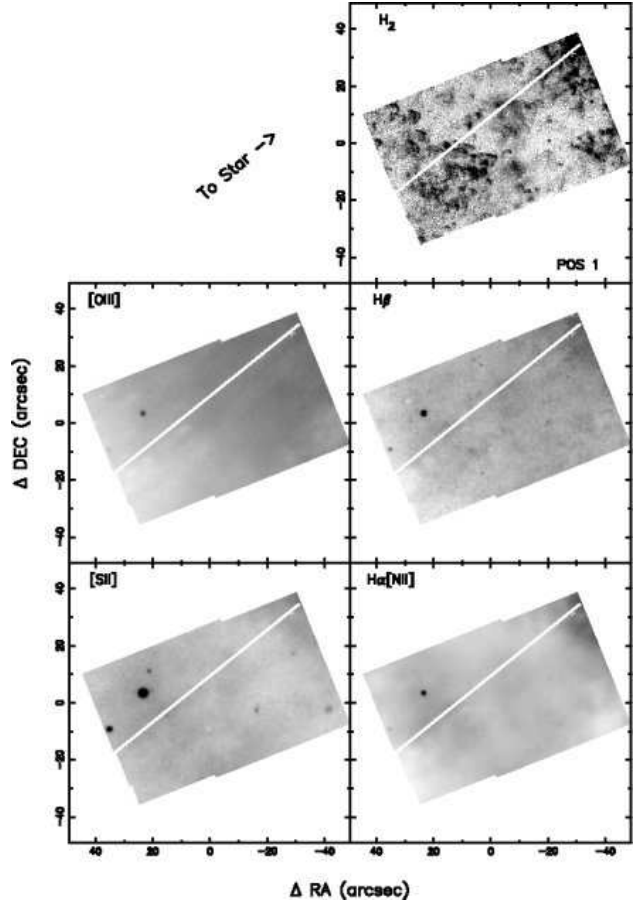}
\caption{NICMOS position~1.   The observed H$_2$ 2.12~\um\  emission is at the upper right column box, with the [OIII]  (top-left), H$\beta$ (right-center), H$\alpha$/[NII] (right bottom), and [SII] (left bottom)  emission from the same region extracted from the CTIO ground based images  \cite{odell04}.   The white cross (+) marks the location where the flux measurements  of Table~2 were taken.  The white line shows the location of the crosscut shown in Fig.~\ref{nicprof}.  The approximate direction of the central star is shown with the label. \label{nicpos1}}
\end{figure}

\begin{figure}
\plotone{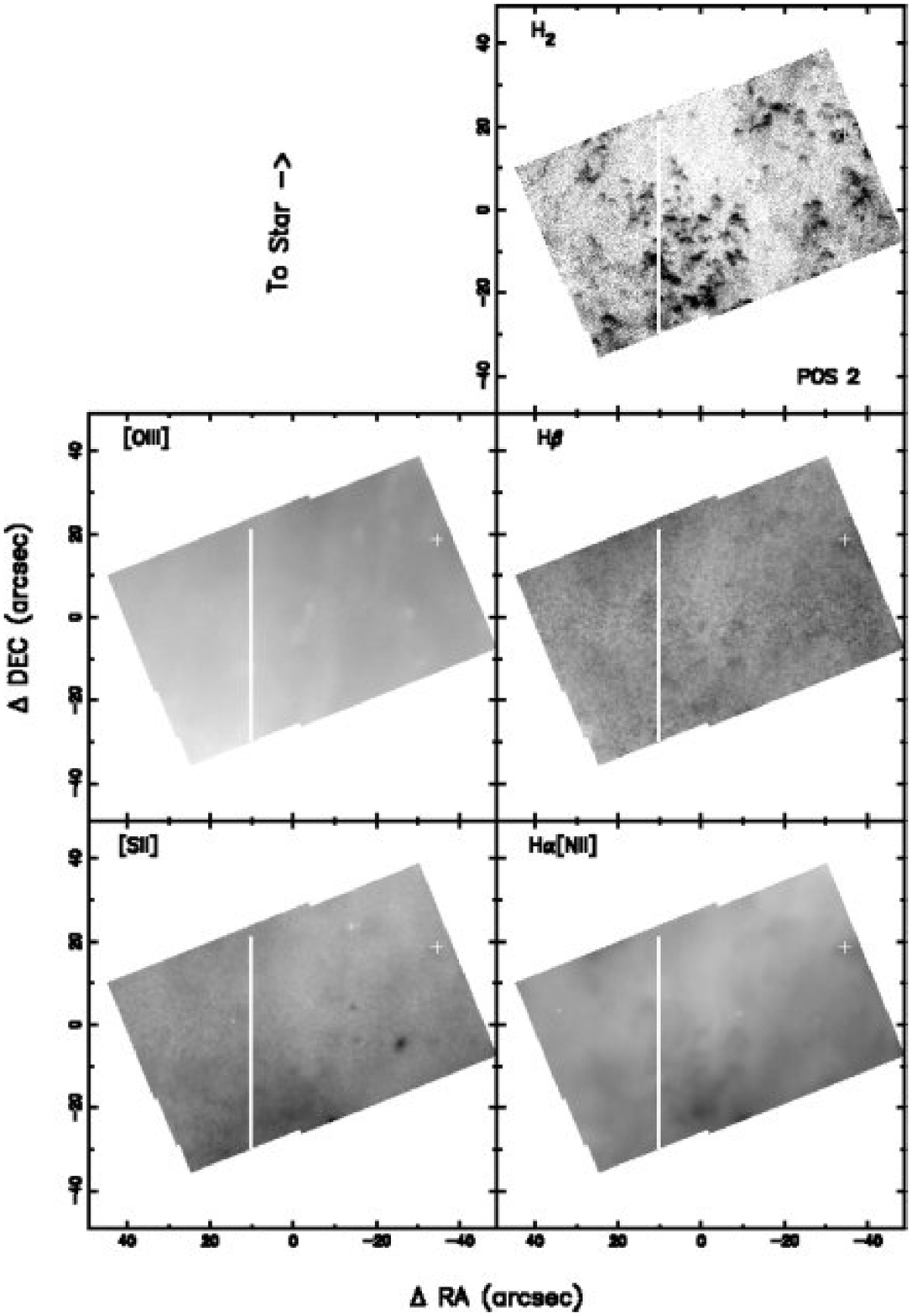}
\caption{NICMOS position 2, same as figure 1.    \label{nicpos2}}
\end{figure}

\begin{figure}
\plotone{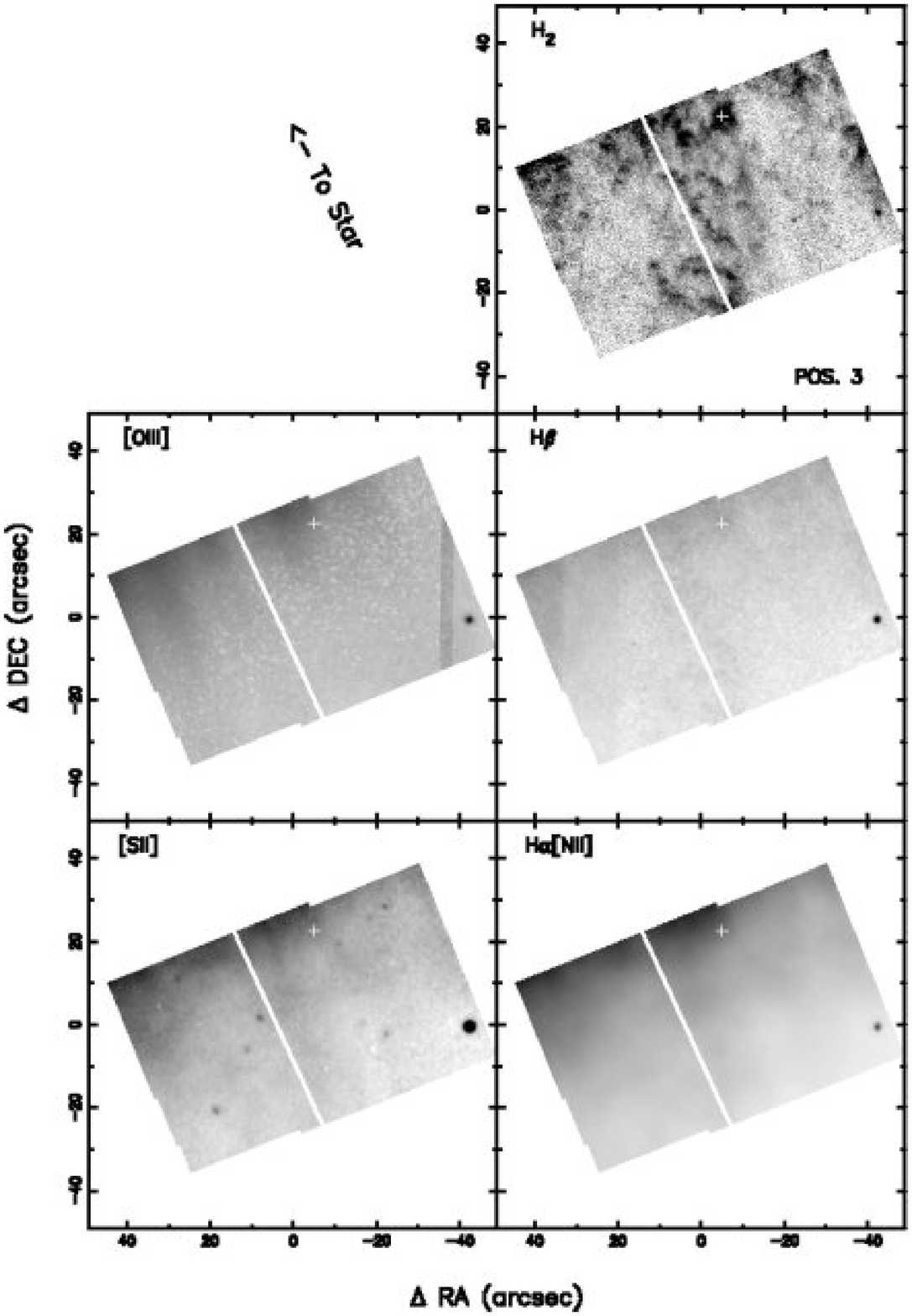}
\caption{NICMOS position 3, same as figure 1.    \label{nicpos3}}
\end{figure}

\begin{figure}
\plotone{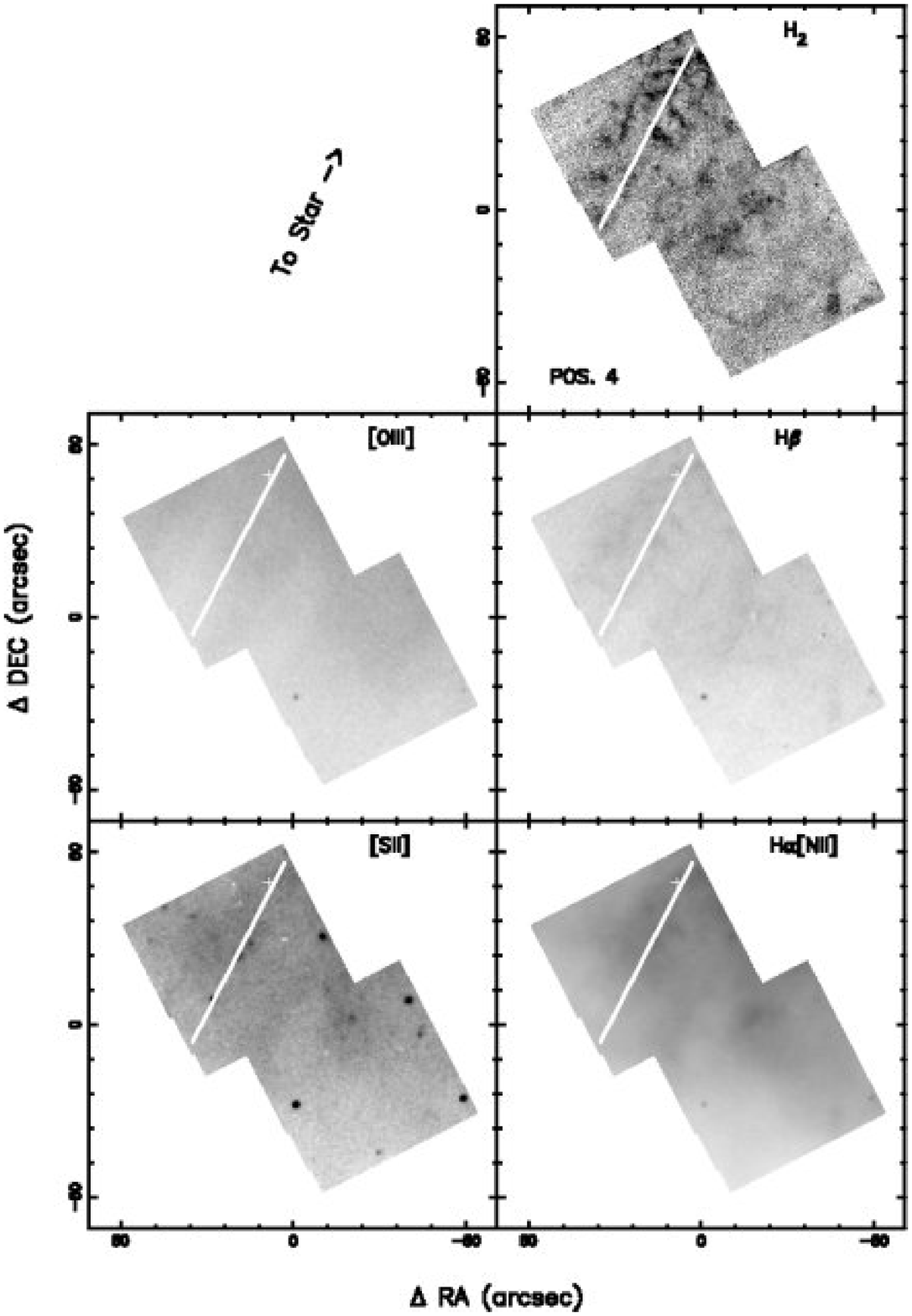}
\caption{NICMOS position 4, same as figure 1.    \label{nicpos4}}
\end{figure}

\begin{figure}
\plotone{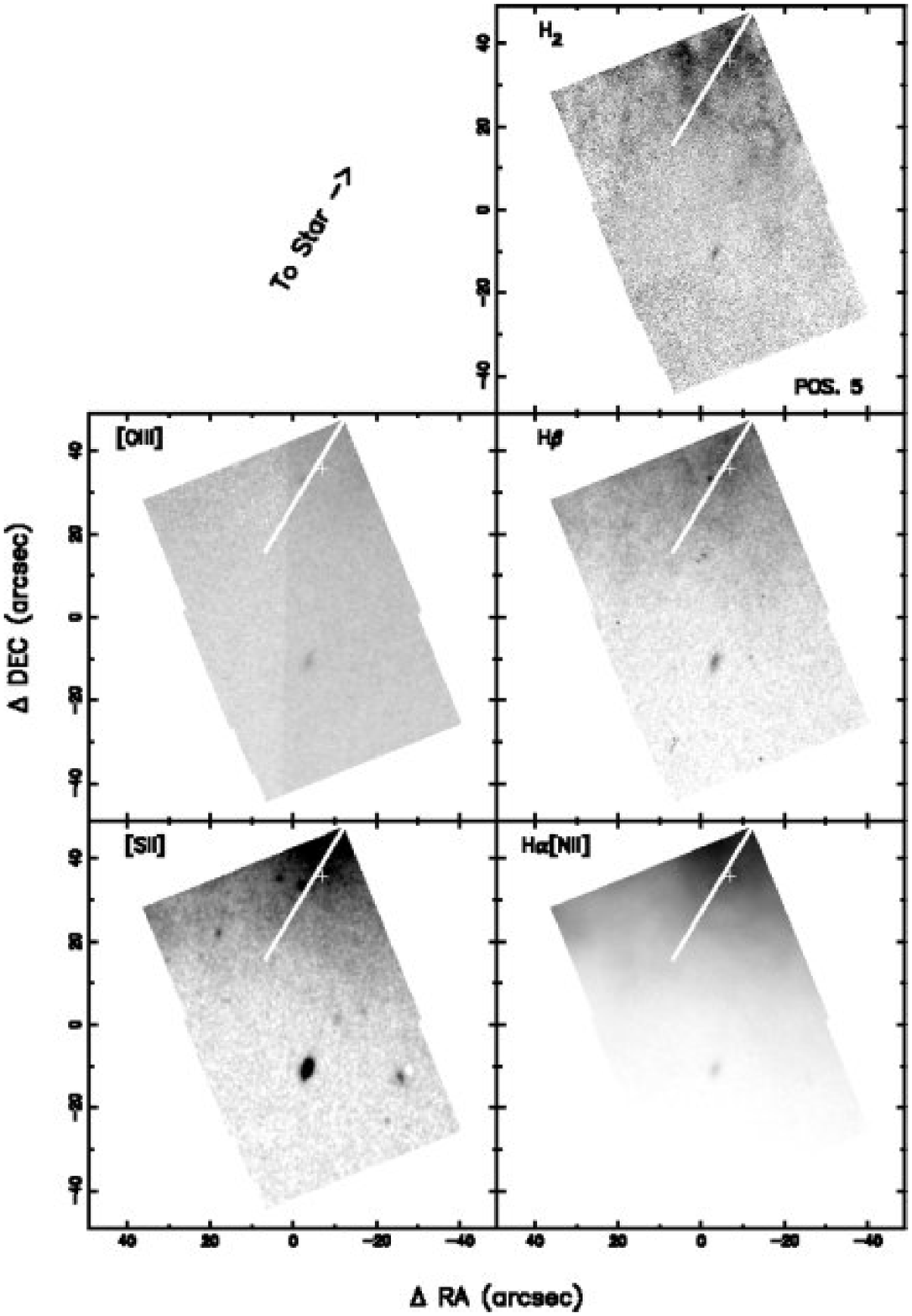}
\caption{NICMOS position 5, same as figure 1.    \label{nicpos5}}
\end{figure}

\begin{figure}
\epsscale{.7}
\plotone{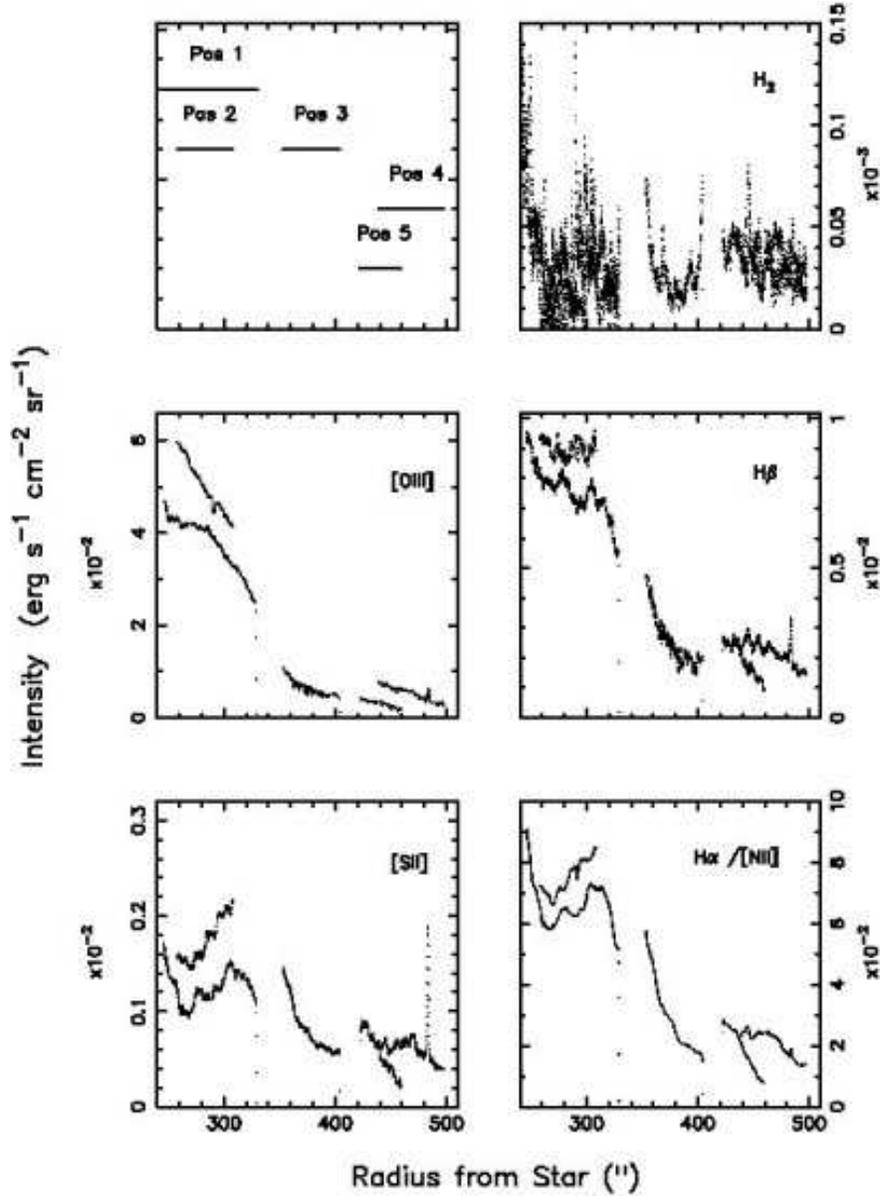}
\caption{The combined intensity radial  profiles of NICMOS positions 1--5 shown as a function of distance from the central star for the different emission line tracers as labeled in the boxes. The top left box  shows the horizontal location of the position slices in relative radial distances from the central star.   Variations in intensity at the same radial distance but different positions is caused by the azimuthal variation of the nebular intensity emission.    \label{nicprof}}
\end{figure}

Figures \ref{nicpos1}, \ref{nicpos2}, \ref{nicpos3}, \ref{nicpos4} and \ref{nicpos5} compare the NICMOS H$_2$ 2.12~\um\ line emission images to the ionized gas tracers of [OIII], H$\beta$, H$\alpha$/[NII] and [SII] from \cite{odell04}  for the respective field positions~1, 2, 3, 4, and~5 .   Our NIC3 images have better sensitivity  and angular resolution than the image in the \cite{speck02} and we observe  H$_2$ line emission at much larger distances than seen in their large scale, ground based mosaic.   The NICMOS field positions~1 and~2 lie closer to the central star at approximately the same projected distance.  The field position~3  follows next in radial distance with positions~5 and~4 overlapping at the farthest radial distances (Fig.~\ref{nicprof}).  In all of the NICMOS field positions the H$_2$  emission is highly  structured revealing arcs and pillars of emission that point towards the central star  (Figures~\ref{nicpos1}--\ref{nicpos5}). This highly structured appearance contrasts with the more smooth, and less structured appearance of  the ionized lines.  This difference indicates that the  H$_2$ line emission is confined to the high density neutral gas of the cometary knots \citep{odell96}.   On the other hand, the ionized gas emission arises from both the more diffuse nebula (50~cm$^{-3}$) and the cometary knots. Closer inspection of all 5~positions reveals that the H$\beta$ emission structures correlates very well with the structures observed in H$_2$.  The [SII] emission  appears to correlate with the H$_2$ emission in positions~1 and~2, but does not show the structure  as well as the H$\beta$.  This suggests that the [SII] emission is more extended, diffuse or less defined by the cometary knots.   The H$\alpha$[NII], which appears like a combination of the H$\beta$ and [SII] emission,  follows the H$_2$ emission but is much more diffuse in appearance than the H$_2$ or H$\beta$.

In positions~1 and~2, the uniformity of the   [OIII] emission  is punctuated by dark shadows of cometary knots that appear clearly in H$_2$ emission.   The  H$_2$ knots must lie in front of some highly ionized gas emitting in order to  cause the extinction of the [OIII]  emission.  The fact that most of the H$_2$ knots do not  appear as [OIII] shadows indicates that most of the H$_2$ knots are confined within the neutral disk and below the highly ionized, diffuse gas apparent in [OIII] emission.  The [OIII] emission trails off in intensity in positions~3 and~4 having a diffuse and streaked appearance with no cometary knot emission. The [OIII] emission at these more distant positions probably arises from above the disk.  The bright streams of emission  in the [OIII] line may be examples of the crepuscular rays found on the large scale [OIII] emission by \cite{odell04}.  

The combined  cross-cuts   (Fig.~\ref{nicprof}) dramatically  show the difference between the H$_2$ emission and optical line ionized gas tracers. The ionized gas tracers  appear to smoothly drop with increasing radius. Differences in the intensity vs. radial distance for these ionized gas tracers are due to the photo-ionization structure of the Helix \citep{odell98, odell04}.  Of all the ionized gas tracers shown,  the H$\beta$ profiles reveal some clumped structure at the 5--10\% level on its basically smooth profile.   In contrast, the H$_2$ emission appears to almost randomly fluctuate with radial distance because of the highly clumped, knot structure of the H$_2$ emission.  The H$_2$ intensity does not appear to decrease significantly with radial distance because the individual knots have a small range of intensities.   Positions that overlap in radial distance, 1~\&~2 and 4~\&~5,  are offset in their optical line emissions because  they are located at very different azimuthal positions and the nebula has a distinct variation with azimuthal angle.  Interestingly, the H$_2$ emission does not appear to have offsets in emission in the overlap regions further supporting that the H$_2$ emission appears  to have an almost constant level with radial distance.

Detailed comparison of line intensities of the brighter knots in each field position also support this contrast between optical ionize gas emission and H$_2$ emission.  Table~\ref{surbrtab} lists  the brightnesses of a small region in each field position shown as a white cross in the Figures~\ref{nicpos1}--\ref{nicpos5}.  These regions were selected as bright rims of the  H$_2$  knots and hence measure the brightest H$_2$ emission in the fields.   Our NICMOS images were taken farther out  in the Helix nebula into regions that were below the detection limit of \cite{speck02}, $\sim10^{-4}$ erg~s$^{-1}$ cm$^{-2}$ sr$^{-1}$, and hence complement their picture.  The H$_2$ emission from the  bright H$_2$ clumps  decreases by a factor of two  between the nearest (pos1) and most distant (pos5) clumps.   In comparison, the ionized line emission drops more steeply with the H$\beta$ emission dropping by a factor of~4 and the [OIII] line dropping by a factor of~5.   

\section{Discussion}

\subsection{ The H$_2$  nebular structure}

\cite{speck02} imaged the entire Helix nebula in H$_2$ showing that it followed the rest of the gas mass tracers and interpreted the H$_2$ as arising in knots in the main disk of the nebula.  However, in their recent  analysis of the ACS and CTIO data in combination with velocity data from the literature,  \cite{odell04} revealed an entirely new 3-D structure of the Helix nebula in which the main ring is broken into an inner disk and an outer ring that are almost perpendicular with respect to each other ($\sim$78\arcdeg).    It is the superposition of these two rings that gives the Helix its helical appearance.  Similar multiple axis structures have been observed easily in edge-on poly-polar planetary nebulae such as NGC~2440 \citep{lopez98}; however, the almost pole-on view of the Helix has made it more difficult to define its geometry.   How does  the full-nebula H$_2$ image of \cite{speck02} fit into this new paradigm for the Helix?

\begin{figure}
\epsscale{.5}
\plotone{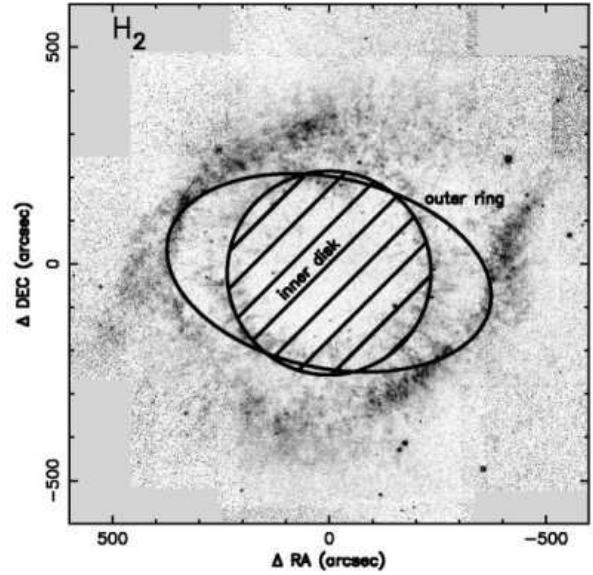}
\caption{H$_2$ line emission image from Speck et~al.\ (2002) with the inner disk and outer ring structures labelled as identified by \cite{odell04}. \label{fullnebh2}}
\end{figure}

Figure \ref{fullnebh2} shows the inner disk and outer ring structures discussed by \cite{odell04} superposed on the  near-IR  H$_2$ image of \cite{speck02}. The H$_2$ near-IR emission of the inner disk is more cleanly separated from the outer ring structure than is found in the optical line tracers  and supports the conclusion that the inner disk is a separate structure than the outer ring  \citep{odell04}.  This cleaner separation occurs because the  H$_2$ emission arises only from the knots and thus there is better contrast than for the optical line tracers that are diluted by the diffuse emission.  The H$_2$ emission arises at the outer  edge of the inner disk that is filled with higher excitation ionized gas, as traced by [OIII], in the center \citep{odell04}. The H$_2$ emission arising in the inner disk is also much fainter than the H$_2$ emission in the outer ring.  The opposite is true for the ionized gas tracers: they are much brighter in the inner disk than the outer ring.  This reversal suggests that the molecular gas in the inner disk has been largely photo-dissociated in comparison to the outer ring and, hence, the cometary knots are more evolved in the inner disk.  

The outer ring does not appear very ring-like in the H$_2$ emission, but more like a ring with southern and northern arc-extensions  at the eastern and western edges that  surround the  northwest and southeast ``plumes" mentioned by \cite{odell04}.  In fact, the H$_2$ emission appears brightest  in these plumes.  The CO kinematics suggest that  these plumes are coherent structures \citep{young99}.   One possible explanation is that the outer ring  defined by \cite{odell04}  is really part of an outer bipolar structure and the southern and northern arcs are the limb-brightened edges of the bicones in the outer structure.  Such an interpretation is in line with the Helix nebula being a poly-polar planetary nebula being viewed pole-on.  The edge-on poly-polar PN, NGC~2440,  has an inner torus of H$_2$ emission and an outer bipolar nebula of H$_2$ emission with axes of symmetry that are tilted with respect to one another \citep{latter95}.

Our NICMOS field positions appear to lie in some of the fainter  or non-existent regions of H$_2$ emission as observed by \cite{speck02} which had less sensitivity than our study.  Positions~1 and~2 are located outside the southern edge of the inner disk in what appears to be a gap between the inner disk and outer structures.  Position ~3 lies to the southwest of the outer structure where no apparent H$_2$ emission appears in the H$_2$ image by \cite{speck02}.   Positions~4 and~5 are located even further away from previously detectable H$_2$ emission south of the nebula.   Positions~3, 4 and~5 also lie in regions where no HI 21~cm line emission was detected by \cite{rodriguez02} nor CO emission detected by \cite{young99}.    The number density  of H$_2$ knots and area filling factor of H$_2$ emission  decreases with increasing radius and is the lowest in positions~3, 4 and~5  (Figs.~\ref{nicpos3}--\ref{nicpos5}).   In the large beams of these radio line observations ($\sim$31\arcsec\ FWHM for CO and  $\sim$42\arcsec\ FWHM for HI 21~cm) , the intensity of the neutral gas emission from these knots is beam diluted and falls below the detection limit of the radio observations.    

\subsection{The multitude of molecular knots in the Helix}

\begin{figure}
\epsscale{.6}
\plotone{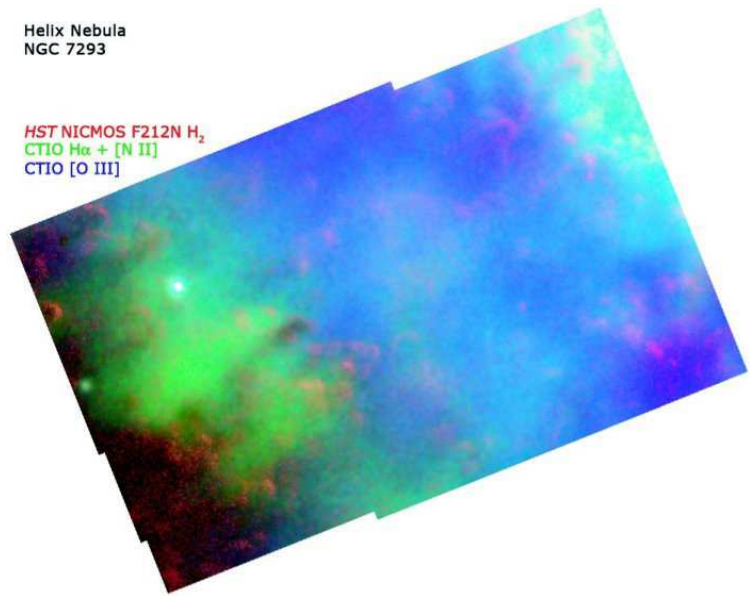}
\caption{NICMOS position 1 in three colors representing the H$_2$, the H$\alpha$+[NII] and the [OIII] as shown in the legend.   This color comparison re-emphasizes the clumpy nature of H$_2$ line emission in comparison to the more diffuse ionized gas line emission. \label{nicpos1_color}}
\end{figure}

Our observations clearly show that the morphology of the molecular hydrogen emission is highly clumped in comparison to the ionized gas tracers.  Figure~\ref{nicpos1_color} shows a multi-color image comparing the [OIII], the H$\alpha$/[NII] and the H$_2$ emission for position~1.  The structure observed in this image is primarily due to the H$_2$ emission  clumps.   In fact,  the H$_2$ emission  images are striking by their lack of diffuse H$_2$ emission.  Close inspection of the more intense regions shows they are composed of overlapping knots of H$_2$ emission.  Hence, we confirm the conclusion of   \cite{speck02} that  the molecular hydrogen is  confined to the high density knots such as seen in the optical by \cite{odell96}.    A similar conclusion was reached by \cite{speck03} for the Ring Nebula based on comparison of high resolution ground-based H$_2$ emission images  to the optical HST images.  Thus in two evolved planetary nebulae, the Helix and the Ring Nebulae,  the H$_2$ line emission is highly structured and confined to  knots.

The near-IR H$_2$ emission provides us with the highest angular resolution map of the neutral gas knots in the Helix nebula.  Previous work has suggested that  the knots contain all the neutral gas detected at substantially lower angular resolution in the CO emission ($\sim$31\arcsec\ FWHM)  by \cite{young99}, in the CI emission ($\sim$15\arcsec\ FWHM) by \cite{young97},  the  HI emission ($\sim$42\arcsec\ FWHM)  by \cite{rodriguez02} and the H$_2$ line emission  ($\sim$4\arcsec\ FWHM)  by \cite{speck02}.  However, all previous neutral gas studies have had insufficient resolution and sensitivity to  separate  and determine the structure  and number density of these neutral gas knots.    The optical study of the knots by \cite{odell96}  provided an initial, lower limit  for the total number of cometary knots  to be 3500 in the entire nebula. They base this estimate by extrapolating the number density of knots they can identify in their WFPC2 optical images vs.\  radius  to the entire nebula. Our H$_2$ images reveal that many more molecular knots exist as defined by the H$_2$ emission arcs than can be identified in the optical images (Figures~\ref{nicpos1}--\ref{nicpos5}).   For example, in the NICMOS field position~1 the number of knots that  appear as [OIII] shadows are less than 10; however,  the number of arc-shaped H$_2$ emission structures is $\sim$150.  In Table~\ref{numdentab}, we list the number of knots, which we identify by arcs of H$_2$ emission,  and  the FOV of the image.  The number density of knots is simply the total divided by the FOV. The area filling factor of H$_2$ emission is the percentage of the FOV that contains H$_2$ emission structures above a ~1$\sigma$ threshold intensity ($\sim$1--$2 \times 10^{-5}$  erg~s$^{-1}$ cm$^{-2}$ sr$^{-1}$).  The total number of knots, the number density  of knots and the area filling factors are the highest for positions~1 and~2, and decrease with larger radial distance from the star as seen in positions~4 and~5.  Interestingly, the peak surface brightness of the  knots does not decrease significantly  with radial distance as we see in Figure~\ref{nicprof} and Table~\ref{surbrtab}.   

We estimate  the total number of molecular knots in the Helix by scaling the number of knots we observe in our NICMOS images to the total angular size of the Helix.   If we look at the H$_2$ image of \cite{speck02} we find that our NICMOS field positions~1 and~2 land in a region of average or slightly below average H$_2$ intensity.   So, we base a conservative estimate of the total number of knots by using the average knot number density of positions~1 and~2, 0.041 knots/arcsec$^2$.  The  H$_2$ emission region is an annulus with an inner radius of $\sim$170\arcsec\ and an outer radius of 450\arcsec\ covering a total angular area of $5.5\times 10^5$ arcsec$^2$.   Multiplying the average knot number density by the total angular area equals $\sim$23,000 molecular hydrogen knots in the Helix nebula or a factor of 6.5 larger than previous estimates based on optical images \citep{odell96}.  The estimated mass of a single  Helix knot is $\sim$$1.5\times 10^{-5}$~M$_\odot$ \citep{odell96} or 10$^{-4}$~M$_\odot$ \citep{young97}.  However, the \cite{young97} CI study defined a ``knot"  to be the size of 
30\arcsec~$\times$10\arcsec\ in size, which would include $\sim$10 H$_2$ knots if we assume the knot densities of postion 1.  Because our H$_2$ knots appear similar in shape and size to the optical knots observed by \cite{odell96}, we adopt their  mass esimate for individual knots.  If, for simplicity, we assume that all H$_2$ knots have similar properties with an average mass of $\sim$$1.5\times10^{-5}$ M$_\odot$  \citep{odell96}, then the total neutral gas mass of the Helix nebula is $\sim$0.35~M$_\odot$.  Our neutral gas mass estimate is substantially larger than the 0.01~M$_\odot$ estimated by \cite{odell96} who underestimated the total number of knots.   It is  also higher than the 0.18~M$_\odot$ estimated from the CO  observations that were corrected for the atomic CI emitting gas by \cite{young99} who have underestimated the mass because the CO observations do not detect all of the molecular gas.   Our estimate of the neutral gas mass is comparable to the ionized gas mass of 0.3~M$_\odot$ \citep{henry99}, and the total gas mass in the nebula is $>$0.65~M$_\odot$  in the main part of the disk. The mass loss rate, that created the main disk over $\sim$28,000 years  is  \.M~$> 2.3\times 10^{-5}$~M$_\odot$ y${-1}$.  Such a large mass loss rate supports the independent conclusion that the Helix's progenitor star was massive,  6.5~M$_\odot$,  resulting in a present day core mass of 0.93~M$_\odot$ \citep{gorny97}.  

\subsection{The  evolution of  H$_2$  knots}

\begin{figure}
\plotone{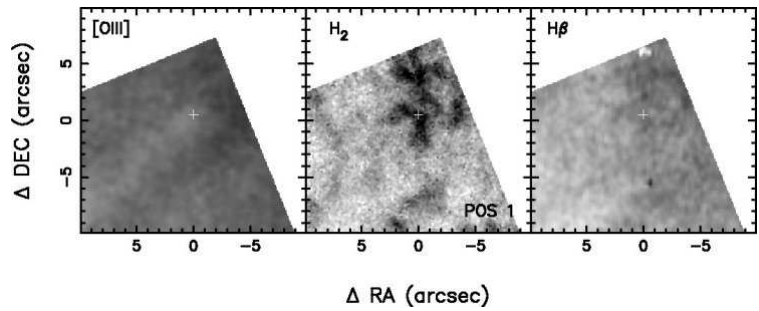}
\caption{The selected H$_2$ knot from NICMOS position 1 shown  in [OIII]  (shadow),  H$_2$ and H$\beta$ line emission.  The white cross, which marks the location of the line emission measurement,  is a guide for the relative position of these emissions.  The H$_2$ and H$\beta$ line emissions appear spatially coincident while the [OIII] shadow column ends at the cross and the H$_2$ arc emission structure.  \label{knot1}}
\end{figure}

Figure~\ref{knot1} shows a close up of the bright knot in NICMOS field position~1 in three tracers, [OIII], H$_2$ and H$\beta$.  The knot extinguishes the [OIII] emission and its long shadow runs off to the bottom left corner.  The arc-shaped head of the knot is clearly observed in the H$_2$ emission and is apparent, at lower contrast, in the H$\beta$ emission.   Close comparison of the molecular hydrogen emission shows that it almost coincides with the H$\beta$ and is displaced towards the star from the shadowed regions of the the [OIII] absorption of the knot (Fig.~\ref{knot1}).   This structure suggests that the H$_2$ emission arises in mini-PDRs on the clump surfaces that point toward the central star. The intensities measured for the H$_2$ emission are consistent with the PN-PDR models of \cite{natta98}.  Comparison of this knot structure  (Fig.~\ref{knot1})  with the detailed study of an optically bright knot in the inner most regions of the Helix \citep{huggins02} reveals a  change in the knot's structure. The most inward knot studied in H$_2$ by \cite{huggins02} is  pillar-like with a well developed crown structure. The H$\alpha$ emission in this knot lies closer to the central star compared to the H$_2$ emission and the H$_2$ emission lies closer to the central star than the CO emission of the knot. This stratified structure is what we expect for a PDR.  However, we do not see a separated stratification of the H$_2$ and H$\beta$ emission from our example knot from Position~1.  This difference in layered vs.\ not layered structure of the knots'  PDRs suggest that the inner most knot has a more evolved PDR front than the Position~1 knot.

A comparison of the different field positions shows a progression in the morphology of the H$_2$ knots with radial distance from the central star.   The positions closest to the central star  (positions~1 and~2) have numerous knots, with cometary structures, i.e. arcs at the top spires.  The farthest positions (4 and~5)  have substantially less knots and less structure to the knots.  Position~3, which lies in between the two extremes,  has a middle density of knots that are lined in  continuous rows with fewer spire structures.    A similar trend from highly structured to almost amorphous was observed in the  H$\alpha$  structure of the knots by \cite{odell04}.  The H$_2$ emission traces this morphological change even further out into the nebula showing that the neutral gas clumps  have even less structure PNs at the outer edge of the nebula.  This progression supports the idea that the initial stages of the structure formation is caused by  instabilities in the interacting winds front  or by the ionization front that are later refined by the photo-excitation.   

\subsection{ The  H$_2$  knots as mini-PDRs}

When molecular hydrogen was first imaged in PNs it was found that their brightnesses were too high for the molecular emission to have originated in photo-dissociation regions (PDRs) and thus attributed to shock excitation \citep{beckwith78,zuckerman88}. However, the PDR models used for these comparisons were designed for interstellar molecular clouds, rather than circumstellar nebulae around rapidly evolving stars. As such these models only included Far-UV photons and failed to include the soft X-ray emission inherent from very hot ($>$100,000~K) white dwarfs. The temperature of the white dwarf also changes as the star evolves during the lifetime of the PN. Furthermore, the original models assumed that the cloud was homogeneous, while the gas around PNs is clearly very clumpy in structure. In addition, the interstellar clouds are not expanding, whereas this is the case for the gas around PNs, causing a change in the optical depth of the gas and therefore allowing the photons to penetrate the gas more easily. The models of \cite{natta98}  included three of these factors (evolving central star, expanding gas, x-ray photons) into their PDR model for PNs, and showed that the molecular hydrogen emission was approximately consistent with excitation of H$_2$ in PDRs in these environments. The  \cite{natta98} model has been applied to three PNs, NGC~2346 (Vicini et~al.\ 1999), the Helix nebula \citep{speck02}, and the Ring Nebula \citep{speck03}. 

Our new observations of the Helix molecular knots confirms that the PDR gas in the Helix resides in a multitude of mini-PDRs, not in a diffuse molecular component \citep{speck02, huggins02}.  The other, perhaps more suprising fact is that the intensity of individual knots remains fairly constant with distance from the central star in the range 5--$9\times10^{-5}$  erg~s$^{-1}$ cm$^{-2}$ sr$^{-1}$ (Table~\ref{surbrtab}; Fig.~\ref{nicprof} ).  The apparent, almost random variation in the H$_2$ line  intensity with respect to radius (Fig.~\ref{nicprof}) occurs because the number density of knots varies along the cross cut not the intensity of the knots themselves.  The relative consistency of the H$_2$ line  intensity of the individual knots with respect to  distance from the central star further supports that the PDRs are distributed as mini-PDRs and there is not one PDR front for the Helix but a multitude of them.  The  higher rotational lines of H$_2$, that were observed with ISOCam by \cite{cox98}, indicate a thermalized temperature of the gas to be $\sim$900~K  that appears to be independent of distance from the central star.   The gas density within the knots, $\sim$10$^4$--10$^5$~cm$^{-3}$ is high enough to thermalize the flourescently excited H$_2$ gas.   The observed  H$_2$ line intensities of the Helix and the derived  molecular gas temperature are approximately consistent with the PN/PDR theory of \cite{natta98}.    For  evolution of the most massive stars,  e.g.\  core mass  0.836~M$_\odot$,  at the age of the Helix, $\sim$16,000 years,  \cite{natta98} predict H$_2$ gas temperatures of  $\sim$1000~K and H$_2$  2.12~$\mu$m line intensities of $\sim$$5\times10^{-5}$  erg~s$^{-1}$ cm$^{-2}$ sr$^{-1}$  in good agreement with the observed temperature  and the individual H$_2$ line intensities of the majority of knots.  At this stage of the PN PDR evolution, the H$_2$ line intensity  decreases only gradually with time and the heating of the molecular gas is dominated by the soft--X-ray emission of its 123,000~K central star 
\citep{bohlin82}.  

Despite this approximate success of the PN/PDR models of \cite{natta98},  these models fall short of complete success.  The brightest  H$_2$  intensity  is almost a factor of 10 larger than the  \cite{natta98} prediction.   Recent model calculations of the H$_2$ intensity of knots in radiative equilibrium with the stellar radiation field by \cite{odell05} also under-predict the H$_2$ line intensities.  The solution to this underprediction of both models,  may be to combine the time evolution aspects of the  \cite{natta98} models with the inclusion of knots, as modelled by \cite{odell05}.  For example,  the time-dependent process of photo-evaporation causes an advection of the H$_2$ from the surface of the knot and a propagation of the PDR front into the knot which may boost the H$_2$ emission because of the constant photodissociation of fresh molecular gas \citep{natta98}.  Evidence for photoevaporation has been found in the ionized gas studies of the inner most knots  \citep{odell00},  which are knots that are directly exposed to the central star light.   Thus, the higher H$_2$ knots could be those that are directly exposed to the central starlight and experiencing photoevaporation.  However, most knots experience a softer starlight that has been attenuated by intervening knots of molecular gas and dust.   Secondly, the  brightest H$_2$ intensity detected by 
\cite{speck02},  $\sim$$3\times10^{-4}$ erg~s$^{-1}$ cm$^{-2}$ sr$^{-1}$, may be the result of multiple knots along the line of sight,  i.e.\ filling factors greater than~1 that have  a multiplying effect on the intensity.  This raises an interesting point that the spatial distribution of H$_2$ over the entire nebula varies not because of substantial H$_2$ intensity variation, as one might expect in a PDR front.  Rather,  the apparent H$_2$ surface brightness is proportional to the number density of H$_2$ knots. 

\section{Conclusions}

New observations of  H$_2$ 2.12~$\mu$m line reveal several new aspects to the molecular knots of the Helix nebula. The H$_2$ images reveal that the knots have arcuate structures with the apex pointing towards the central star.   These molecular hydrogen knots are most highly structured in the field positions closest to the central star and become increasingly less structured with increasing radius. All of the H$_2$ emission is confined to knots.  In contrast the ionized gas tracers have a significant component of diffuse ionized gas emission. Using the number density of molecular hydrogen knots in the 5~NICMOS  field positions, we estimate the total number of knots to be $\sim$23,000, a factor of 6.5 more than previous estimates based on  optical images.  The total neutral gas mass in the Helix based on these new knots estimates is 0.35~M$_\odot$ assuming an average mass of $\sim$$1.5\times10^{-5}$~M$_\odot$  for the individual knots based on previous work by \cite{odell96}. The H$_2$ emission structure of  the entire Helix nebula  supports the recent interpretation of the Helix as a nearly  pole-on  poly-polar planetary nebula.  The average intensity is 5--$9\times10^{-5}$ erg~s$^{-1}$ cm$^{-2}$ sr$^{-1}$ remains relatively constant with projected distance from the central star.  The temperature and H$_2$ 2.12~$\mu$m  intensity of the knots suggest an origin in the photodissociation  regions (PDRs) of the individual knots; however, theoretical models for  the PDRs in planetary nebulae do not adequately reproduce the  H$_2$ intensity.  The brightest knots appear in regions of more numerous knots and may be exposed to direct starlight that may cause rapid photoevaporation in comparison to the more embedded knots of the disk.

\acknowledgments
We gratefully acknowledge the work of many STScI colleagues who contributed to this observational setup of this project. Zoltan Levay who superposed the NICMOS fields on the combined CTIO and ACS image.    This work was supported in part by an STScI grant  GO 01041  and by the internal STScI funds, DDRF D0001.82319.

\begin{table}
\begin{center}
\caption{ NICMOS field position centers and the H$_2$ knot positions with respect to these centers  \label{loctab}}
\begin{tabular}{lcccc}
\tableline\tableline
Positions & RA(2000) of (0,0)  & Dec(2000) of (0,0) & $\Delta$RA to H$_2$ knot &   $\Delta$Dec to  H$_2$ knot \\
 &  (h~ m~ s)\phn\phn & (\arcdeg~ ~\arcmin~ ~\arcsec)~ & (\arcsec) & (\arcsec) \\
\tableline
central star & 22 29 38.5\phn &$-$20 50 13.5\phn \\
pos1		&22 29 53.03	 &$-$20 53 27.87      &$-$28.32	     &31.57\\
pos2		&22 29 40.23  &$-$20 54 44.07      &$-$34.51       &18.68\\
pos3		&22 29 27.44  &$-$20 56 \phn0.26 &\phn$-$4.97  &22.53\\
pos4		&22 29 59.67  &$-$20 56 21.79	    &\phn\phs7.11 &41.01\\
pos5		&22 29 44.98	 &$-$20 58 10.38      &\phn$-$7.01   &35.72\\
\tableline
\end{tabular}
\end{center}
\end{table}

\begin{table}
\begin{center}
\caption{Surface Brightnesses of Field Positions  \label{surbrtab}}
\begin{tabular}{lccccccc}
\tableline\tableline
Positions & Radius  &  H$_2$ &   H$\alpha$/[NII] &  H$\beta$ &  [OIII] (5007 \AA) & [SII] (6725 \AA) \\
of   H$_2$ knot & (\arcsec) &  \multicolumn{5}{c}{ ( erg s$^{-1}$ cm$^{-2}$ sr$^{-1}$)} \\
\tableline 
pos1		&	253	&   1.27365E-04	& 7.95596E-02	 & 9.33381E-03  & 4.31994E-02 &  1.42868E-03\\
pos2		&	257        &    9.38642E-05	& 8.19865E-02 & 1.05037E-02 &  6.45470E-02 &  1.51075E-03 \\
pos3		&	360	&   8.55222E-05 &	4.51389E-02 & 9.29548E-03 & 9.49064E-03 & 1.09471E-03 \\
pos4		&	448        & 8.61214E-05	& 2.47975E-02 & 2.90198E-03 & 6.68796E-03 & 6.28393E-04 \\
pos5		&	436	&  5.99320E-05	 & 2.42383E-02  &  2.36069E-03 & 3.43582E-03 & 6.77509E-04 \\
\tableline
\end{tabular}
\end{center}
\end{table}

\begin{table}
\begin{center}
\caption{Number of H$_2$ Knots  \label{numdentab}}
\begin{tabular}{lcccc}
\tableline\tableline
Positions & Total & FOV area & number density &  area filling factor \\
                  &            &                  &                               &        of H$_2$ emission \\
\tableline
pos1		&	150 & $50\arcsec\times\phn80\arcsec$ & 135 knots/arcmin$^2$ & 80\% \\
pos2		&  180 & $50\arcsec\times\phn80\arcsec$ & 162 knots/arcmin$^2$ &  70\%  \\
pos3		&  \phn90 & 	$50\arcsec\times\phn80\arcsec$ & \phn81 knots/arcmin$^2$ &  60\%   \\
pos4		&	\phn90 & $50\arcsec\times100\arcsec$ & \phn65 knots/arcmin$^2$ &  50\%  \\
pos5		&	\phn20 & $50\arcsec\times\phn80\arcsec$ & \phn18 knots/arcmin$^2$ &  25\%  \\
\tableline
\end{tabular}
\end{center}
\end{table}

\begin{thebibliography}{}

\bibitem[Beckwith et~al.(1978)]{beckwith78} Beckwith, S., Gatley, 
I., \& Persson, S.~E. 1978, \apjl, 219, L33 

\bibitem[Bohlin et~al.(1982)]{bohlin82} Bohlin, R.~C., 
Harrington, J.~P., \& Stecher, T.~P. 1982, \apj, 252, 635 

\bibitem[Burkert \& O'dell(1998)]{burkert98} Burkert, A., \& 
O'dell, C.~R. 1998, \apj, 503, 792 

\bibitem[Capriotti(1973)]{capriotti73} Capriotti, E.~R. 1973, 
\apj, 179, 495 

\bibitem[Cox et~al.(1998)]{cox98} Cox, P., et~al. 1998, 
\apjl, 495, L23 

\bibitem[Gon{\c c}alves et~al.(2001)]{goncalves01} Gon{\c c}alves, 
D.~R., Corradi, R.~L.~M., \& Mampaso, A.  2001, \apj, 547, 302 

\bibitem[Gorny et~al.(1997)]{gorny97} Gorny, S.~K., Stasinska, 
G., \& Tylenda, R. 1997, \aap, 318, 256 

\bibitem[Harris et~al.(1997)]{harris97} Harris, H.~C., Dahn, 
C.~C., Monet, D.~G., \& Pier, J.~R. 1997, IAU Symp.~180: Planetary 
Nebulae, 180, 40 

\bibitem[Henry et~al.(1999)]{henry99} Henry, R.~B.~C., Kwitter, 
K.~B., \& Dufour, R.~J. 1999, \apj, 517, 782 

\bibitem[Huggins et~al.(2002)]{huggins02} Huggins, P.~J., 
Forveille, T., Bachiller, R., Cox, P., Ageorges, N., \& Walsh, J.~R. 2002, 
\apjl, 573, L55 

\bibitem[Huggins \& Healy(1986)]{huggins86} Huggins, P.~J., \& 
Healy, A.~P. 1986, \apjl, 305, L29 

\bibitem[Kastner et~al.(1994)]{kastner94} Kastner, J.~H., Gatley, 
I., Merrill, K.~M., Probst, R., \& Weintraub, D. 1994, \apj, 421, 600 

\bibitem[Latter et~al.(1995)]{latter95} Latter, W.~B., Kelly, 
D.~M., Hora, J.~L., \& Deutsch, L.~K. 1995, \apjs, 100, 159 

\bibitem[Lopez et~al.(1998)]{lopez98} Lopez, J.~A., Meaburn, 
J., Bryce, M., \& Holloway, A.~J. 1998, \apj, 493, 803 

\bibitem[McCullough \& Hubble Helix Team(2002)]{mccullough02} 
McCullough, P.~R., \& Hubble Helix Team 2002, American Astronomical Society 
Meeting Abstracts, 201

\bibitem[Meaburn et~al.(1992)]{meaburn92} Meaburn, J., Walsh, 
J.~R., Clegg, R.~E.~S., Walton, N.~A., Taylor, D., \& Berry, D.~S.\ 1992, 
\mnras, 255, 177 

\bibitem[Meixner et~al.(2004)]{meixner04} Meixner, M., 
McCullough, P., Hartman, J., O'dell, R., \& Speck, A.~K. 2004, 
Astronomical Society of the Pacific Conference Series, 313, 234 

\bibitem[Natta \& Hollenbach(1998)]{natta98} Natta, A., \& 
Hollenbach, D. 1998, \aap, 337, 517 

\bibitem[O'Dell \& Handron(1996)]{odell96} O'dell, C.~R., \& 
Handron, K.~D. 1996, \aj, 111, 1630 

\bibitem[O'dell(1998)]{odell98} O'dell, C.~R. 1998, \aj, 116, 
1346 

\bibitem[O'Dell et~al.(2000)]{odell00} O'Dell, C.~R., Henney, 
W.~J., \& Burkert, A. 2000, \aj, 119, 2910 

\bibitem[O'Dell et~al.(2002)]{odell02} O'Dell, C.~R., Balick, 
B., Hajian, A.~R., Henney, W.~J., \& Burkert, A. 2002, \aj, 123, 3329

\bibitem[O'Dell et~al.(2004)]{odell04} O'Dell, C.~R., 
McCullough, P.~R., \& Meixner, M. 2004, \aj, 128, 2339 

\bibitem[O'Dell et~al.(2005)]{odell05} O'Dell, C.~R., 
Henney, W.~J., \& Ferland, G.~J. 2005, \aj, in press. 

\bibitem[Rodr{\'{\i}}guez et~al.(2002)]{rodriguez02} 
Rodr{\'{\i}}guez, L.~F., Goss, W.~M., \& Williams, R. 2002, \apj, 574, 179 

\bibitem[Speck et~al.(2002)]{speck02} Speck, A.~K., Meixner, 
M., Fong, D., McCullough, P.~R., Moser, D.~E., \& Ueta, T. 2002, \aj, 123, 
346 

\bibitem[Speck et~al.(2003)]{speck03} Speck, A.~K., Meixner, 
M., Jacoby, G.~H., \& Knezek, P.~M. 2003, \pasp, 115, 170

\bibitem[Thompson et~al.(1998)]{thompson98} Thompson, R.~I., 
Rieke, M., Schneider, G., Hines, D.~C., \& Corbin, M.~R. 1998, \apjl, 492, 
L95 

\bibitem[Vicini et~al.(1999)]{vicini99} Vicini, B., Natta, A., 
Marconi, A., Testi, L., Hollenbach, D., \& Draine, B.~T. 1999, \aap, 342, 
823 

\bibitem[Young et~al.(1999)]{young99} Young, K., Cox, P., 
Huggins, P.~J., Forveille, T., \& Bachiller, R. 1999, \apj, 522, 387 

\bibitem[Young et~al.(1997)]{young97} Young, K., Cox, P., 
Huggins, P.~J., Forveille, T., \& Bachiller, R. 1997, \apjl, 482, L101 

\bibitem[Zanstra(1955)]{zanstra55} Zanstra, H. 1955, Vistas in 
Astronomy, 1, 256 

\bibitem[Zuckerman \& Gatley(1988)]{zuckerman88} Zuckerman, B., \& 
Gatley, I. 1988, \apj, 324, 501 

\end{thebibliography}
\end{document}